\documentclass[a4paper,11pt]{article}
\pdfoutput=1 

\usepackage{jinstpub} 

\title{\boldmath Free ion yield of Trimethyl Bismuth used as sensitive medium for high-energy photon detection}

\author[a,1]{M.~Farrad{\`e}che\note{Corresponding author.},}
\author[a]{G.~Tauzin,}
\author[a]{J-Ph.~Mols,}
\author[b]{J-P.~Dognon,}
\author[c]{V.~Dauvois,}
\author[a]{V.~Sharyy,}
\author[a]{J-P.~Bard,}
\author[a]{X.~Mancardi,}
\author[a]{P.~Verrecchia,}
\author[a]{and D.~Yvon}

\affiliation[a]{CEA, DRF, IRFU, Universit\'e Paris-Saclay, F-91191 Gif-sur-Yvette, France}
\affiliation[b]{CEA, DRF, NIMBE, CNRS, Universit\'e Paris-Saclay, F-91191 Gif-sur-Yvette, France}
\affiliation[c]{CEA, DEN, DANS, DPC, SECR, LRMO, Universit\'e Paris-Saclay, F-91191 Gif-sur-Yvette, France}

\emailAdd{morgane.farradeche@cea.fr}

\abstract{The CaLIPSO project is an innovative high-energy photon detector concept using trimethylbismuth as sensitive medium in a liquid ionization chamber. The detector, designed for high precision brain PET imaging, works as a time-projection chamber and detects Cherenkov light and charge signal.
We measured the free ion yield of trimethylbismuth, which represents the number of electron-ion pairs released by the incident photon. To do so, we developed a low-noise measuring system to determine the current induced by a \Co~source in the liquid with an accuracy better than 5 fA for an electric field up to 7 kV/cm. We used tetramethylsilane as benchmark liquid to validate the apparatus and we measured a zero-field free ion yield of $0.53 \pm 0.03$ in agreement with measurements in literature.
However, we found a zero-field free ion yield of $0.083 \pm 0.003$ for trimethylbismuth, which is a factor 7 lower than the typical values for similar dielectric liquids.
Quantum chemistry computations on heavy atoms tend to demonstrate a high capacity of trimethylbismuth to capture electrons which could explain the weak value. The consequences of a low free ion yield in terms of high-energy photon detection and brain PET imaging are finally discussed.}


\newcommand{\dd}{\mathrm{d}}
\DeclareMathOperator{\e}{e}
\newcommand{\cube}{\ensuremath{^3}}		
\newcommand{\squared}{\ensuremath{^2}}	
\newcommand{\minusone}{\ensuremath{^{-1}}}	
\newcommand{\Gfi}{\ensuremath{G_{\mbox{\textit{fi}}}}}
\newcommand{\ro}{\ensuremath{r_{\text{c}}}}
\newcommand{\Co}{\ensuremath{^{60}}Co}

\usepackage{tabularx}

\usepackage{subfig}
\usepackage{comment}
\usepackage{xcolor}

\begin{document}
\maketitle
\flushbottom

\section{Introduction} \label{sec:intro}

	The CaLIPSO project is a high-energy photon detector concept designed for precise brain PET imaging. In order to reach a good detection efficiency, liquid trimethyl bismuth (TMBi) is used as sensitive medium in an ionization chamber. Indeed, bismuth is a heavy element ($Z=83$) and TMBi is a high density liquid (2.3 g/cm\cube), thereby maximizing the photoelectric cross section. The high energy photon is thus efficiently converted in a single primary electron.
This electron ionizes the detection medium and releases electron-ion pairs drifted in a strong electric field. This ionization signal allows energy measurement and 2D interaction positioning via the pixelated detector.

\begin{figure}[htbp]
\centering
\includegraphics[width=0.85\textwidth]{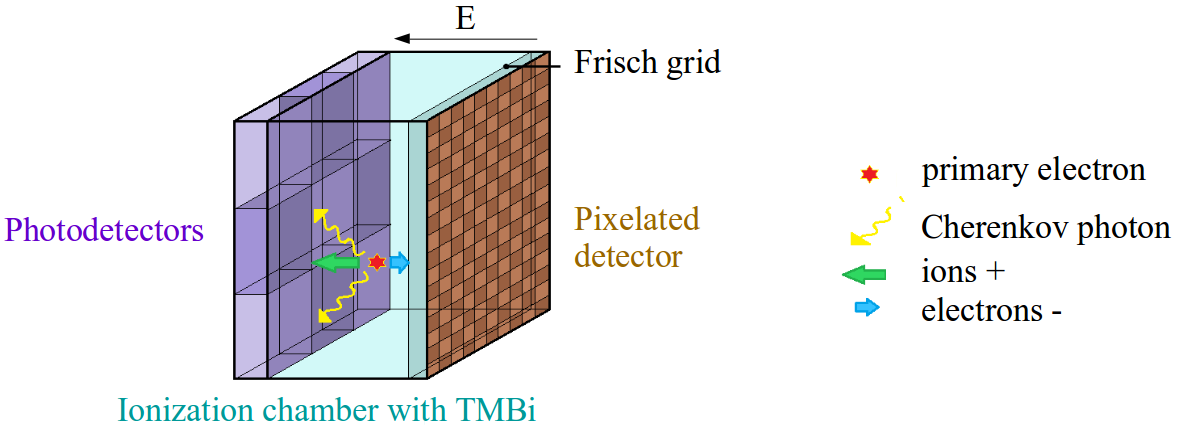}
\caption{The CaLIPSO detector design. Liquid trimethyl bismuth allows a double detection : Cherenkov radiation and ionization signal via the charge drift in an electric field.}
\label{fig_calipso}
\end{figure}

	The primary electron also produces Cherenkov radiation. The development of high-energy photon detection using Cherenkov light gives promising results and is outlined in detail in refs.~\cite{Ram16} and \cite{Can17}.
The CaLIPSO detector works as a time-projection chamber and detects both ionization and light signals as illustrated in figure~\ref{fig_calipso}. The principle of detection is fully described in ref.~\cite{Yvo14}. The simultaneous double detection leads to promising detector performances with a precision on photon conversion positioning in the detector down to 1~mm\cube\ (due to time projection of ionization signal) and a coincidence resolution time better than 150~ps.
	The work presented in this paper focuses on the ionization signal.

\section{Ionization in liquid media}

	\subsection{Onsager theory}
 
	When an electron is released from an atom, it reaches thermal equilibrium by inelastic collisions with neighbouring molecules. If this occurs far enough from its parent ion, so that the Coulomb attraction is low relative to the thermal energy of the medium, the electron diffuses in the liquid. It is then likely to recombine with a positive ion from another ionization. This process is called volume or general recombination.

	In liquids, the electron undergoes many collisions and generally reaches thermal equilibrium in the Coulomb field of its positive parent. The Coulomb attraction predominates and the electron may recombine with its parent ion, resulting in an initial recombination. The critical distance at which thermal energy equals Coulomb energy is given by : $\ro = e^2 / 4\pi \epsilon k_BT$ where $e$ is the elementary charge, $\epsilon$ the liquid permittivity, $k_B$ the Boltzmann constant and $T$ the operating temperature. In dielectric liquids, $\ro$ is typically 230 to 300~\AA.

	Onsager treated the problem of the Brownian movement of an electron under the influence of an additional electric field and the Coulomb attraction of its parent ion only~\cite{Ons38}. This theory can thus be applied when the distance between consecutive ionizations is large when compared to the initial separation distance (i.e. thermalization distance). In tetramethyl silane (TMSi) for example, the mean thermalization length has been measured to be $r=166$~\AA~\cite{Hol91}, while the mean distance of ionization processes is about 3000~\AA\ in hydrocarbon liquids~\cite{Eng96}.

	A distinctive property of Onsager theory is that the first expansion term of the escape probability is independent of the thermalization distance $r$. Then for moderate electric field, the escape probability rises linearly with the external electric field~$E$ as~: 

\begin{subequations}
\begin{equation}
\Pi(E,T) = \e^{-\ro/r} \ ( 1+ \alpha E +...)
\end{equation}
and the slope-to-intercept ratio is predicted as a function of liquid permittivity $\epsilon$ as~:
\begin{equation}\label{eq_alpha}
\alpha \ = \frac{e^3}{8 \pi \epsilon (k_BT)^2 } = \frac{e}{2 k_B T} \ \ro 
\end{equation}
\end{subequations}

	The fraction of electrons released by the primary electron which can be collected is usually expressed by the free ion yield \Gfi\ defined as the number of electron-ion pairs escaping initial recombination per 100~eV of absorbed energy in the detection medium. Within the Onsager theory framework, the free ion yield can be expressed as a function of the escape probability averaged over thermalization distances as :
\begin{equation}
\Gfi(E,T) = N^\text{tot} \int_{0}^{\infty} \Pi(E,T) f(r) \ \dd r
\end{equation}
where $N^\text{tot}$ is the total number of electrons released (before initial recombination) per 100 eV deposited energy and $f(r)$ is the distribution function of electron thermalization distances. Finally, in the linear region described by Onsager theory, we can write :
\begin{equation}\label{eq_onsager}
\Gfi = \Gfi^0 \ (1+ \alpha E)
\end{equation}
where the intercept $\Gfi^0$ is known as the zero-field free ion yield.

	\subsection{Properties of dielectric liquids}

	TMBi has never been studied in the context of particle detection. However, a number of other dielectric liquids have been studied as active media for ionization detectors~\cite{Hol91,Mun92}. We report in table~\ref{table_gfi} the free ion yields and mobilities for a collection of similar dielectric liquids reported in the literature.
	In order to validate our free ion yield measuring apparatus, we first implemented it with a well-documented liquid : tetramethyl silane (TMSi). 

\begin{table}[htbp]
\centering
\caption{Zero-field free ion yields and mobilities at room temperature.}
\label{table_gfi}
\smallskip
\begin{tabular}{|ll|c|c|c|c|}
\hline
                            &      & Permittivity & Density     & Free ion yield $^\text{a}$ & Mobility $^\text{b}$ \\
                            &      & $\epsilon_r$   & $d$         & $\Gfi^0$                   & $\mu$ \\
                            &      &              & (g/cm\cube) &                            & (cm\squared/V/s) \\
\hline
Tetramethyl silane          & TMSi & 1.92         & 0.645       & 0.65 & 90 \\
Tetramethyl germane         & TMGe & 2.01         & 1.006       & 0.63 & 90 \\
Tetramethyl tin             & TMSn & 2.25         & 1.31        & 0.64 & 70 \\
Trimethyl bismuth           & TMBi &              & 2.3         &      & \\
\hline
\multicolumn{6}{l}{$^\text{a}$ Values from ref.~\cite{Hol91} $\qquad$ $^\text{b}$ Values from ref.~\cite{Sch77}}
\end{tabular}
\end{table}

\section{Free ion yield measurement}

	\subsection{Definition}

	If volume recombination is negligible, the number of electrons released per second is proportional to the ionization current $I$ induced by a radioactive source. From this assumption, the free ion yield can be expressed as~:
\begin{equation}\label{eq_gfi}
\Gfi = \frac{I}{e \ \Delta \epsilon}
\end{equation}
where $e$ is the elementary charge and $\Delta \epsilon$ the energy deposited in the medium per unit of 100 eV$\cdot$s\minusone.

	\subsection{Measuring device}
	
	The detector consists in a parallel-plate ionization chamber with an active volume of 1.6~cm\cube\ and a gap between the stainless steel electrodes of 12~mm. A \Co\ radioactive source with an activity of 0.7~MBq is placed at approximately 50~mm from the center of the active volume of liquid. Charges produced by radiations are drifted by an external electric field up to 7~kV$\cdot$cm\minusone and induce a current.

		\paragraph{Low-noise current measurements.}
		
\begin{figure}[htbp]
\centering
\includegraphics[width=\textwidth]{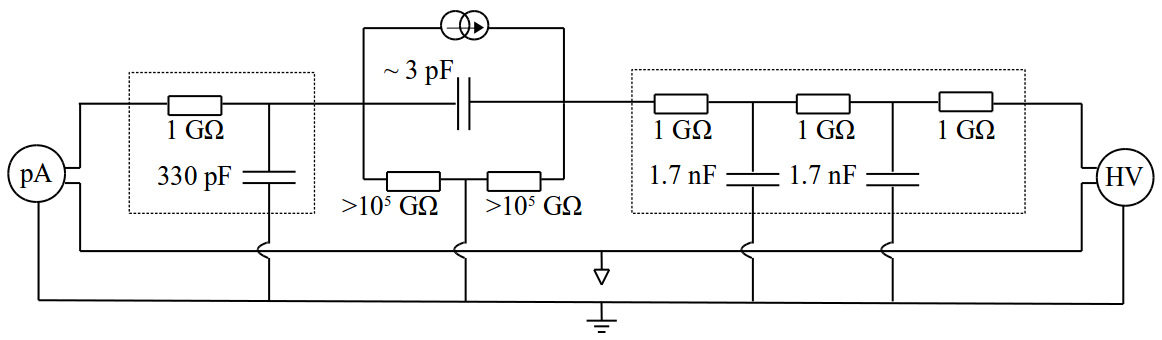}
\caption{Electrical measuring circuit. The ionization chamber is represented by a capacitance of $\sim 3$ pF and the photon-induced current by a current source. The filter placed at the electrometer input protects it from any discharge occurring in the chamber. The high voltage filter damps the noise and ripple from the power supply module.}
\label{fig_circuit_diagram}
\end{figure}

	The ionization current is of the order of a few 10 fA, measured by an electrometer Keithley 6517~B. Negative high voltage is applied with a CAEN N470 module. The electrical circuit diagram is shown on figure~\ref{fig_circuit_diagram}. The radiation-induced current is represented by a current source and the ionization chamber by a capacitance of approximately 3~pF. A leakage current flows in the insulators composing the chamber under the influence of an electric field. We selected Al$_2$O$_3$ ceramic for its chemical compatibility with TMSi and TMBi and its high resistivity ($10^{15}~\Omega\cdot$cm at room temperature~\cite{Sha16}). Considering the geometry of the ionization chamber, the insulation resistance of ceramics is estimated at $10^{14}~\Omega$, corresponding to a leakage current of 70 fA for an operating voltage of 7~kV. It is thereby necessary to divert a large proportion of these currents to the ground as shown on figure~\ref{fig_circuit_diagram}.

	Achieving accuracy and reproducibility at low current and high electric field levels is a technological challenge. The ionization chamber is placed inside an aluminium cage to prevent low-frequency capacitive interferences. A power inverter followed by a screened isolation transformer and a local ground stabilizes the power supply and isolates the device from public power grids. In these conditions, the current induced by the \Co\ source can be measured with an accuracy better than 5~fA for an electric field up to 7~kV$\cdot$cm\minusone.

		\paragraph{Correction from parasitic currents.}
		
\begin{figure}[htbp]
\centering 
\includegraphics[width=0.6\textwidth]{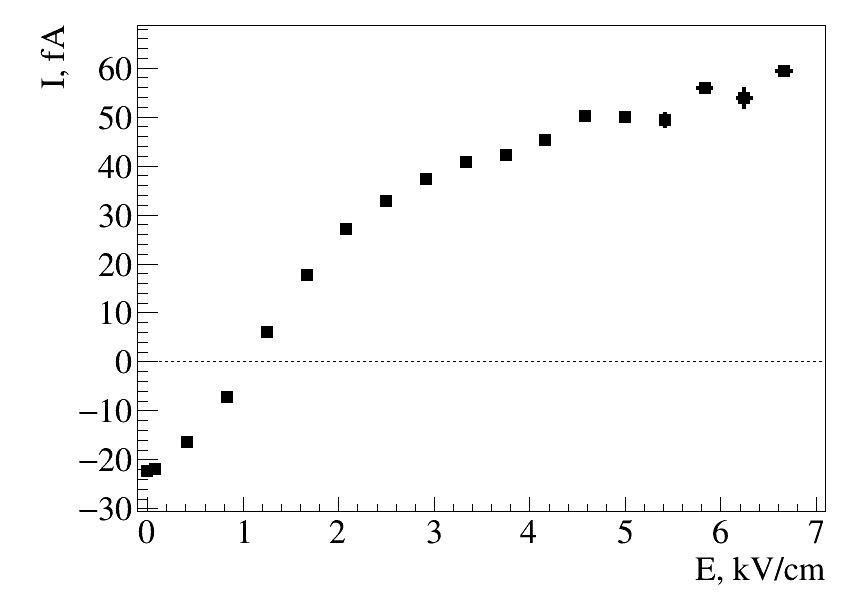}
\caption{Typical measurement of the current induced by the \Co\ source in the ionization chamber under vacuum ($< 1 \cdot 10^{-6}$ mbar) against electric field. Error bars are mostly smaller than markers.}
\label{fig_vacuum} 
\end{figure}

	We observe a non-zero current in the empty chamber ($< 1 \cdot 10^{-6}$ mbar) although the current due to ionization is negligible under vacuum. On figure~\ref{fig_vacuum}, this current is represented as a function of the electric field. We interpret this current as a result of ionizations occurring in wires and in materials composing the chamber. In addition, one can notice on figure~\ref{fig_vacuum} that the zero-field current has an offset induced by Keithley electrometer.
	Thus, in order to assess the ionization current produced by radiation within the liquid, we subtracted the offset and the current measured in the chamber under vacuum from the total current at each biasing voltage.

	\subsection{Monte Carlo simulation}

\begin{figure}[htbp]
\centering
\subfloat[TMSi]{\includegraphics[width=0.49\textwidth]{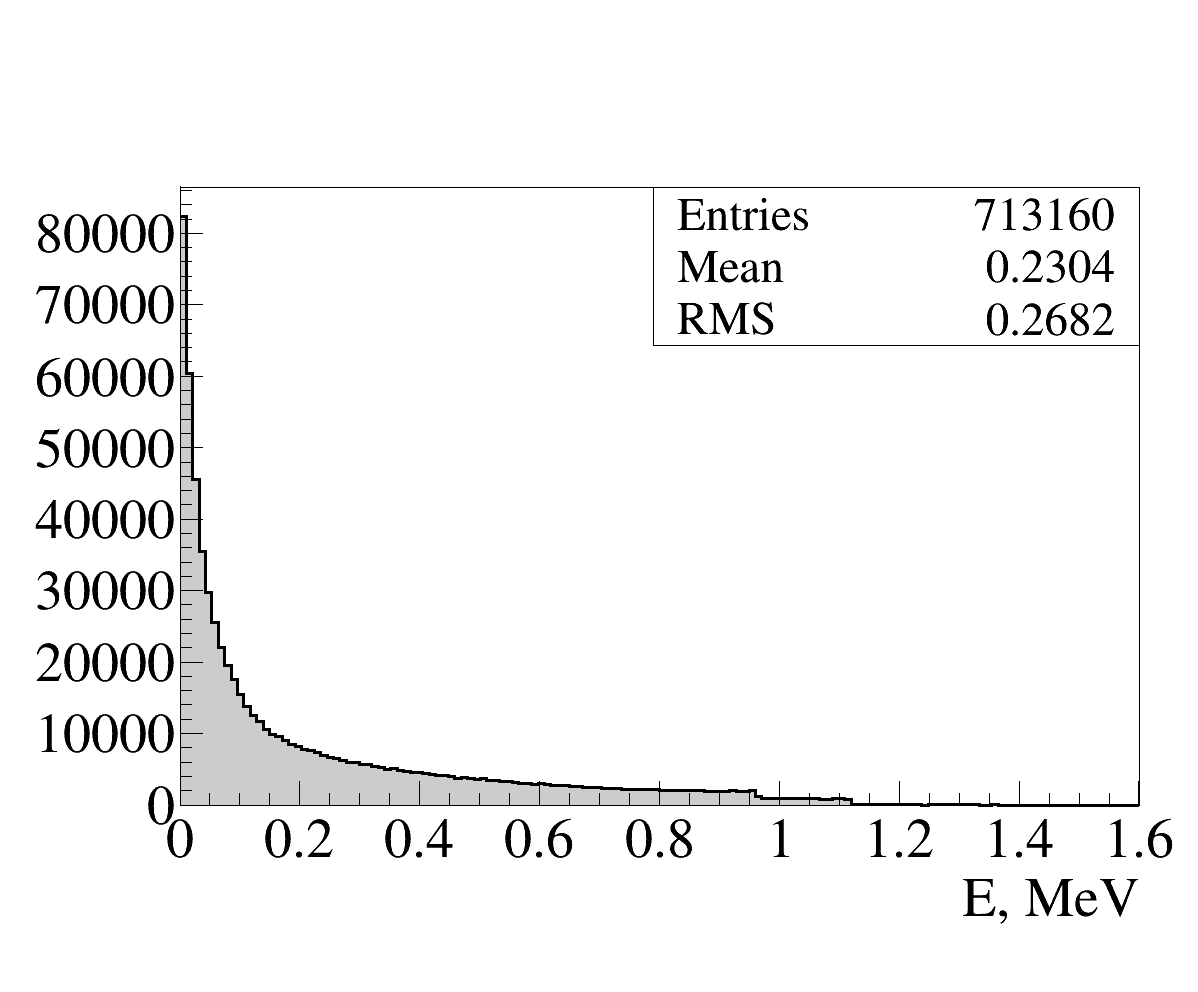}
\label{fig_edepTMSi}}
\hfill
\subfloat[TMBi]{\includegraphics[width=0.49\textwidth]{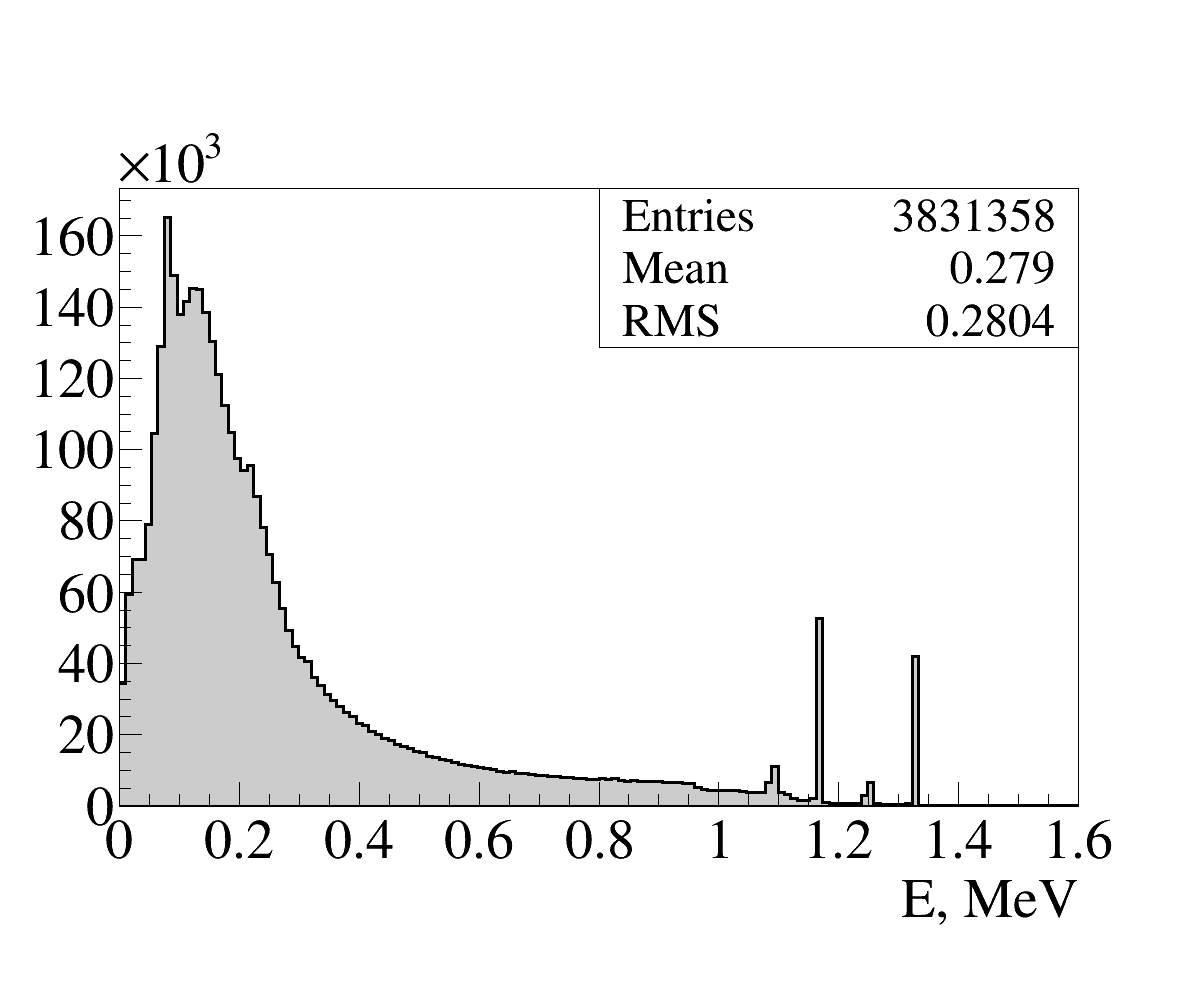}
\label{fig_edepTMBi}}
\caption{Spectra of energy absorbed in the liquids for $10^9$ decays of \Co\ source computed by Monte Carlo simulation.}
\label{fig_edep}
\end{figure}

\begin{figure}[htbp]
\centering
\subfloat[TMSi]{\includegraphics[width=0.41\textwidth]{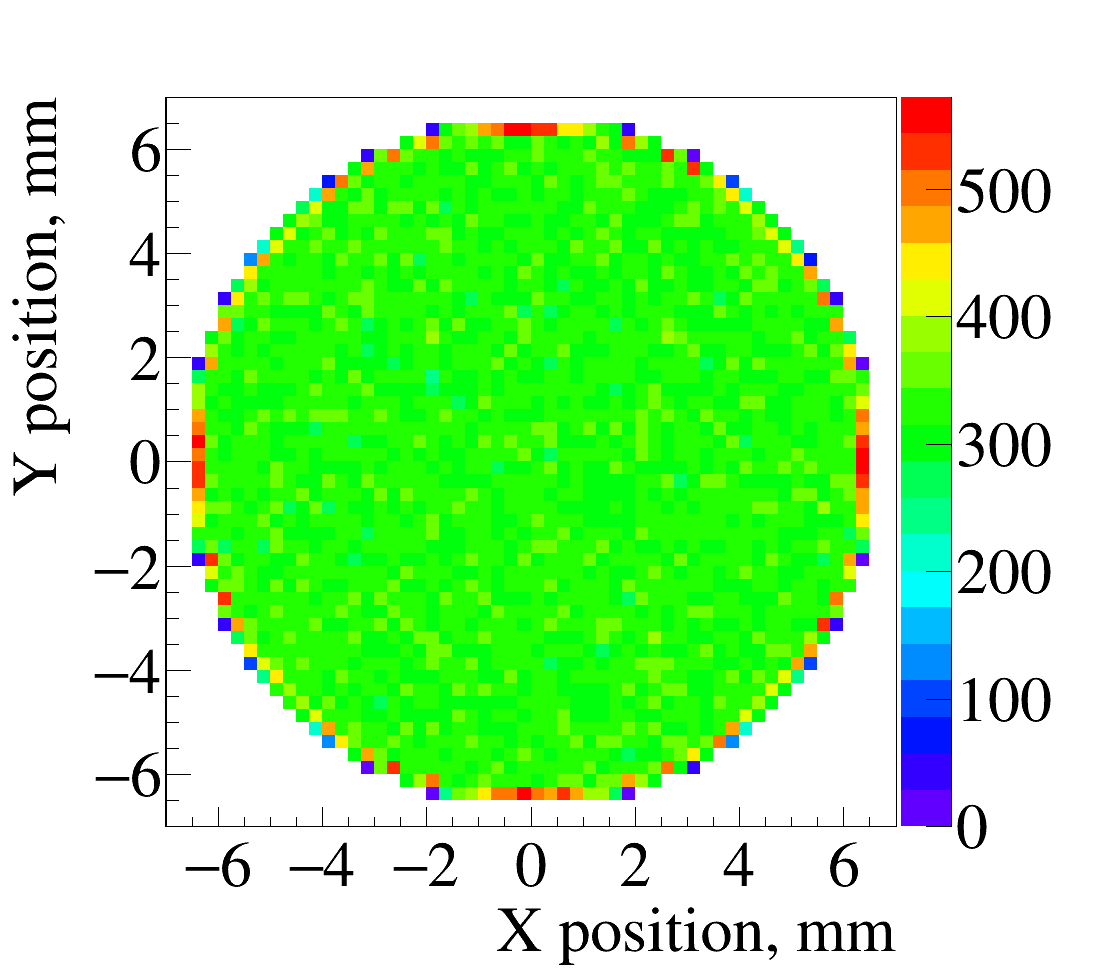}
\label{fig_locdepTMSi}}
\hspace{0.9cm}
\subfloat[TMBi]{\includegraphics[width=0.41\textwidth]{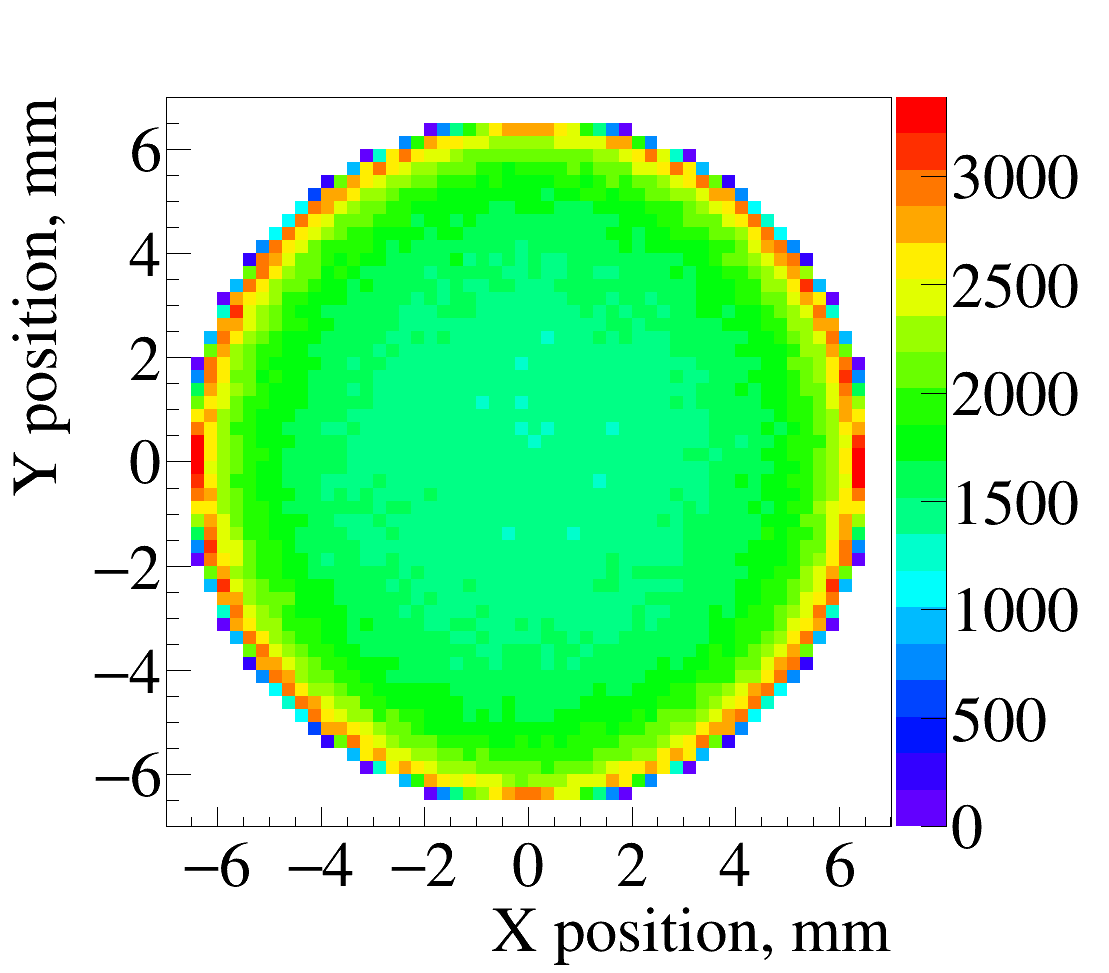}
\label{fig_locdepTMBi}}
\\
\subfloat[TMSi]{\includegraphics[width=0.41\textwidth]{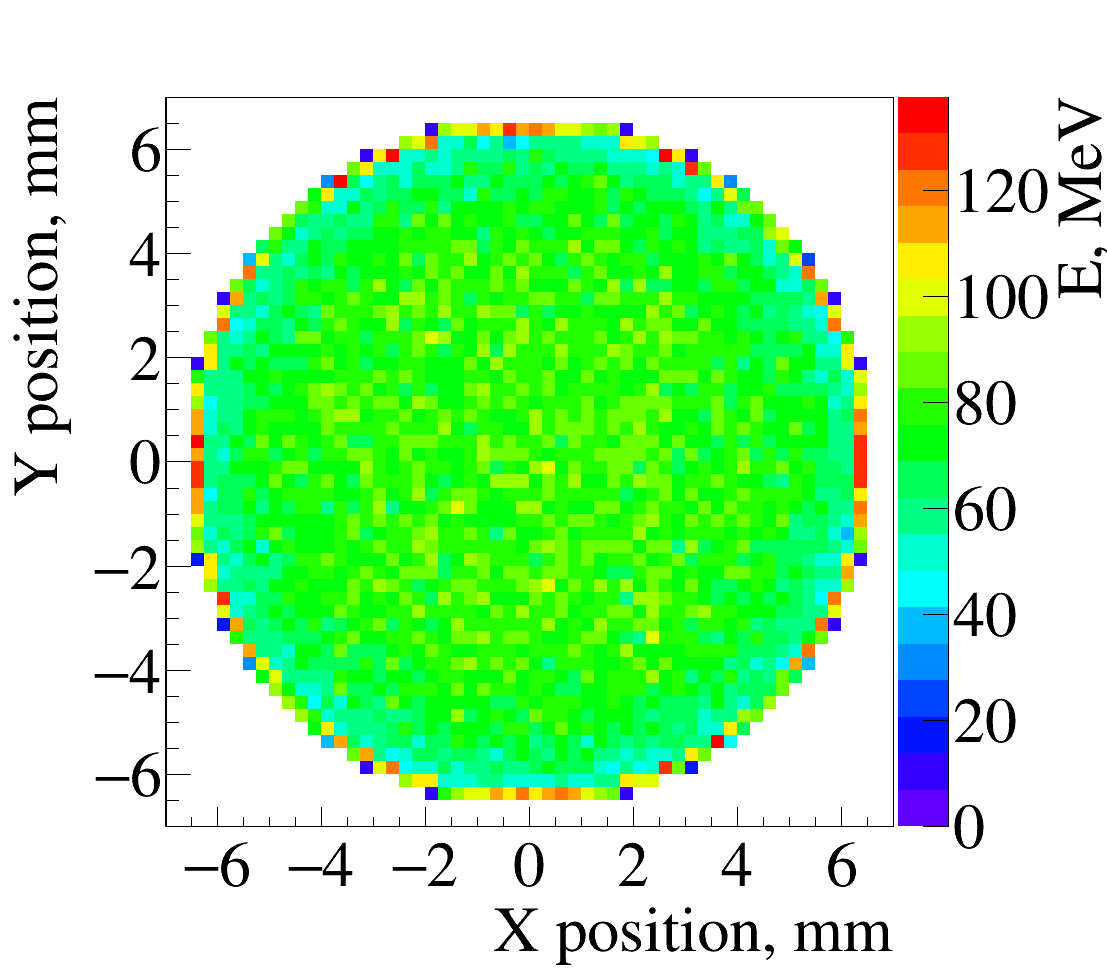}
\label{fig_locdepEdepTMSi}}
\hspace{0.9cm}
\subfloat[TMBi]{\includegraphics[width=0.41\textwidth]{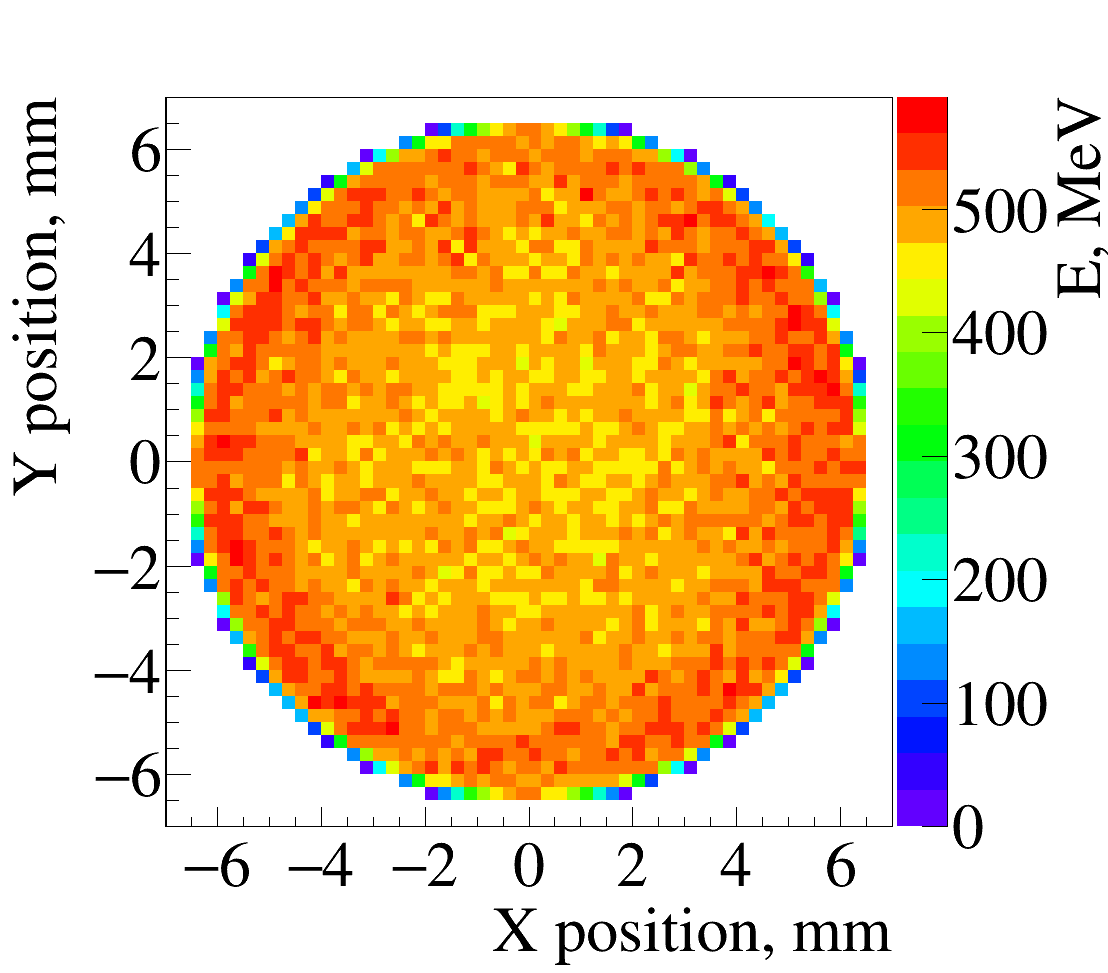}
\label{fig_locdepEdepTMBi}}
\caption{\label{fig_locdep} 2D distributions of interaction positions (a, b) and energy absorbed (c, d) in the liquids for $10^9$ decays of \Co\ source simulated by Monte Carlo simulation.}
\end{figure}

	The ionization chamber and surrounding environment were simulated with a GEANT4 Monte Carlo simulation~\cite{Ago03} within GATE framework~\cite{Jan04}. We computed the spectra of absorbed energy in the liquids (figure~\ref{fig_edep}), and 2D distributions of interaction positions and energy depositions (figures~\ref{fig_locdep}).

	According to simulation, a significant portion of particles depositing energy in TMSi and in TMBi first interacts within the high density ceramics composing the ionization chamber mostly by Compton effect. These scatter photons carry less energy because of their continuous energy spectrum, resulting in a higher probability of interaction in the liquids, noticeable on figure~\ref{fig_edep} as a broad peak at low energies.
2D distributions of interaction positions and energy depositions shown on figure~\ref{fig_locdep} also demonstrate this behaviour. Indeed in both liquids, Compton electrons interact near the edges of the volume and deposit less energy. Moreover in TMBi, the shorter mean free path of high energy photons implies an additional excess of interaction rate near the edges. This energy deposition is still moderate due to the low energy of scatter photons.

\section{Purification}\label{sec_purif}

	A signal loss can be significant due to the presence of electron attachment to impurities. If free electrons are trapped by electronegative molecules, they form anions which are collected on a millisecond time scale, much larger than the free electron collection time. Those anions might recombine with surrounding cations and induce an effect on the ionization current for low electric field strength. Hence, we investigated liquid purification issues in this work. These protocols will be mandatory in the near future, when we will develop ionization chambers with a Frisch grid where only free electrons will contribute to signal formation.

	Ionization chamber and purification system components are cleaned before assembly according to protocols used for ultra-high vacuum technologies. They are immersed in an ultrasonic alkaline bath and rinsed with desalinated water before being dried with filtered argon gas and baked out under vacuum at a temperature up to 250$^\circ$C. Assembly is carried out under a laminar flow wearing clean-room equipment. In these conditions, we reached a vacuum pressure of $1 \cdot 10^{-7}$~mbar stable after the pumping is stopped.

\begin{figure}[htbp]
\centering 
\subfloat[]{\includegraphics[width=0.75\textwidth]{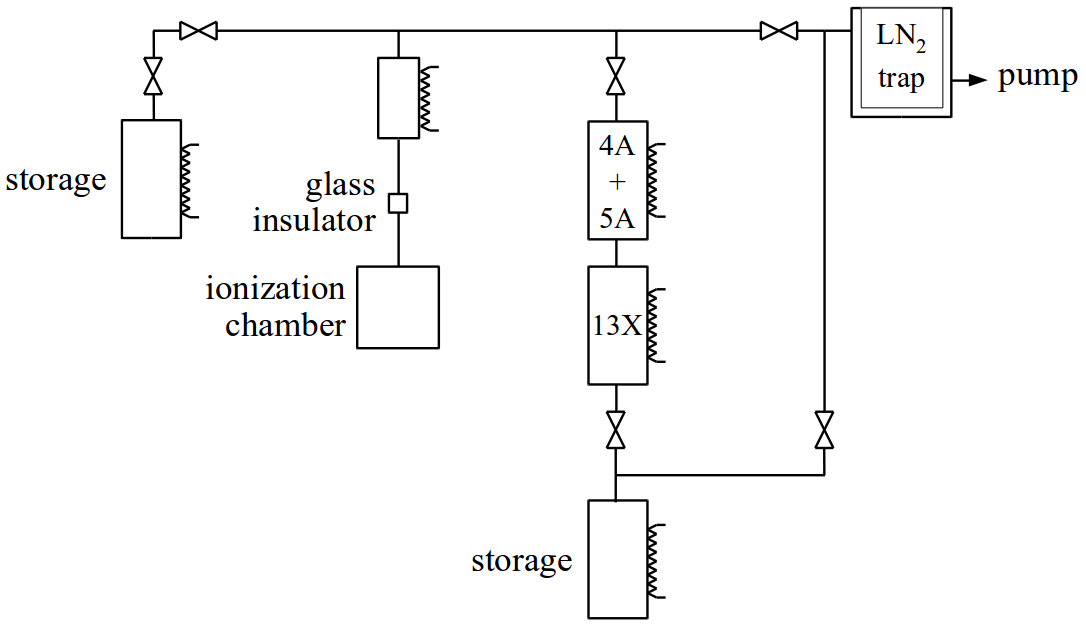}
\label{fig_bench_schema}}
\hfill
\subfloat[]{\includegraphics[width=0.23\textwidth]{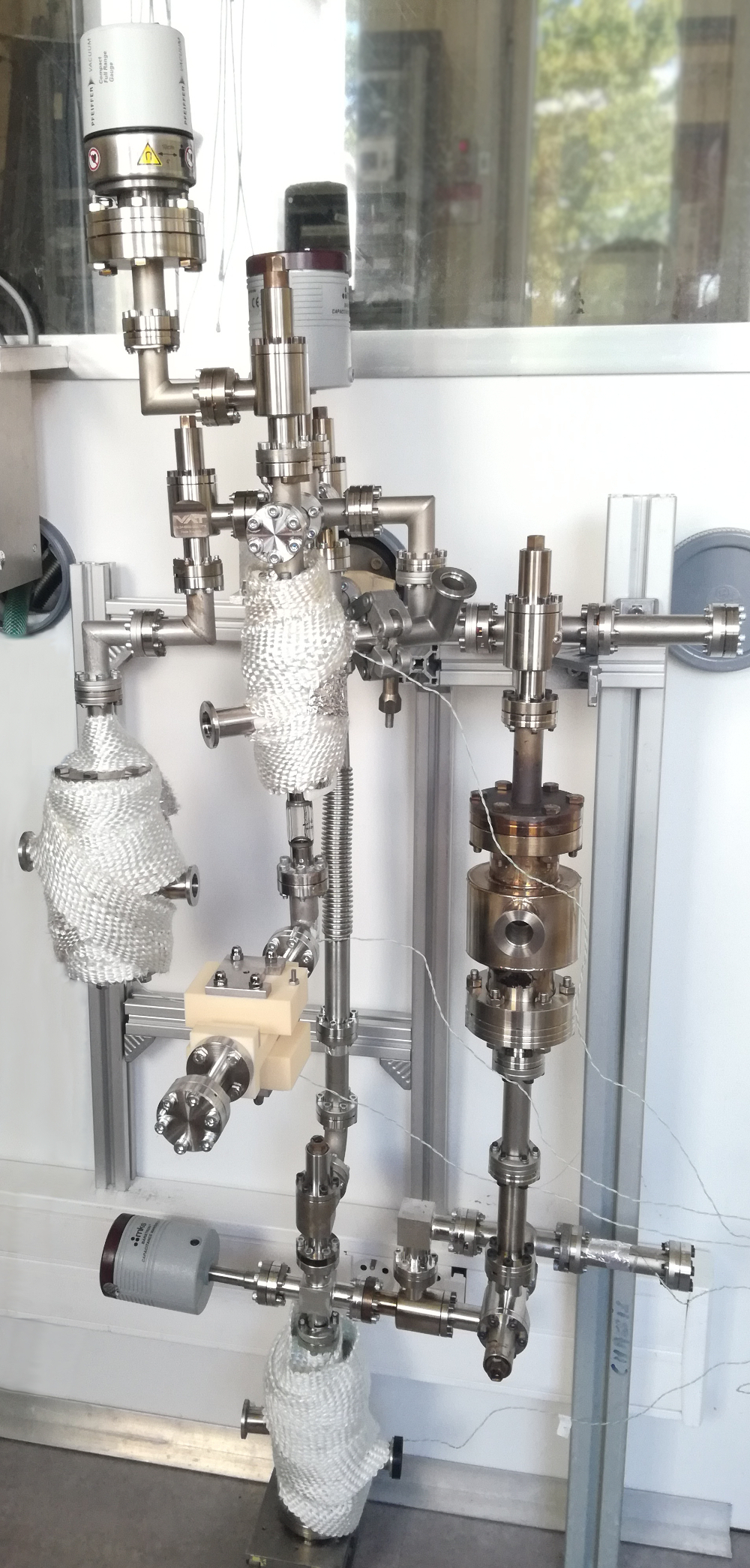}
\label{fig_bench_purif}}
\caption{ Scheme of the purification system (a). Polar molecules of small diameters are removed from the liquid by 4A and 5A molecular sieves and larger molecules by 13X molecular sieve. In the case of TMBi, as shown on photography (b), only 4A molecular sieve is used (see section~\ref{sec_resultats_purif}).}
\label{fig_bench}
\end{figure}

	TMSi provided by Alfa Aesar is supplied with a guaranteed purity of 99.99 \%, and TMBi with a level of impurity inferior to 5 ppm from the supplier JSC-Alkyl. Our purification system was built based on the protocol written by Engler et~al. in ref.~\cite{Eng99} for the experiment KASCADE. The liquids were first degassed to remove protective nitrogen, argon or any other gas dissolved in the liquids. They were then treated in a purification system as illustrated in figure~\ref{fig_bench}. The liquids in vapor phase were forced through molecular sieves, where polar molecules of small diameters are trapped by 4A and 5A molecular sieves and larger molecules by 13X molecular sieve.

\section{Results}

	\subsection{Purification}\label{sec_resultats_purif}

\begin{figure}[htbp]
\centering 
\includegraphics[width=0.75\textwidth]{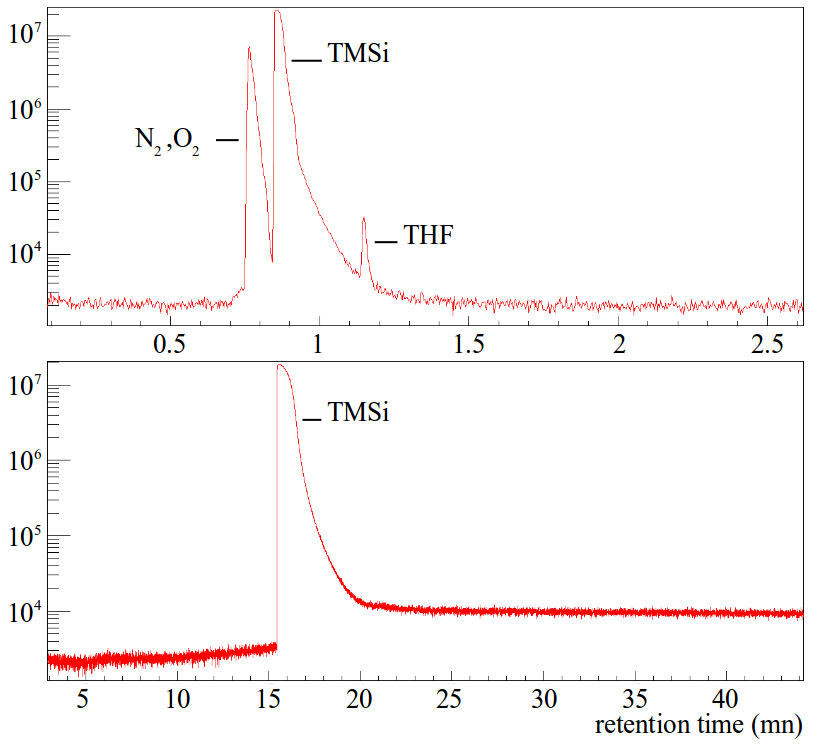}
\caption{GC/MS chromatograms (Total Ion Current TIC) of TMSi samples before (top) and after (bottom) purification. Two different chromatography columns were used with AGILENT analytical device. The peak corresponding to tetrahydrofurane (THF) is no longer detectable after purification. The N$_2$, O$_2$ peak is due to sampling contamination during collection.}
\label{fig_gcms}
\end{figure}

	A preliminary gas chromatography\,/\,mass spectrometry (GC\,/\,MS) analysis was performed with TMSi before purification and revealed the presence of tetrahydrofurane (THF), a residue of TMSi synthesis. After purification, a second analysis highlighted the absence of detectable THF (figure~\ref{fig_gcms}).
In the case of TMBi, the liquid is supplied with ``electronic grade'', which means that no impurities could be noticeable with this type of analysis.
	
	Although TMBi is stable with strongly bounded ceramics such as alumina and silicate glass, we observed a strong reaction with activated alumina and 5A and 13X molecular sieves, whose micro-crystalline structures are based on silicate and alumina. No reaction was observed with 3A and 4A molecular sieves. We deduced that if the sieve pore size is large enough so that TMBi molecules can penetrate these highly polar cages, a reaction occurs between TMBi and sieve compounds. We thus purified TMBi using only 4A molecular sieve in powder form.

	No significant differences were observed in the measurement of the current induced by the \Co\ source before and after purification for either TMSi or TMBi.

	\subsection{Free ion yield}

	The measurement of the ionization current induced by the source in TMSi and TMBi with respect to the applied voltage provides repeatable and reproducible results over several months. The errors were estimated as the variance obtained from the measurement repeated 10 times, combined to the accuracy of the electrometer according to data sheet.

	We deduced the mean energy absorbed per second in the liquids from the energy spectra and event rates computed by Monte Carlo simulation. Systematic uncertainty estimation takes into account uncertainties related to~: (i) the height of active volume depending on tin wire gaskets thickness after compression, (ii) the radius of active volume due to engineering tolerance of manufactured parts of the ionization chamber, (iii) the radioactive source positioning. In addition, owing to the geometry of the ionization chamber, electric field lines in the active volume of liquid are slightly diverted. The signal created in these regions, representing approximately 5 \% of volume, may be loss. According to the simulation, the mean energy deposited in these regions is $5.0 \pm 0.2$~MeV/s in TMSi and $32.8 \pm 0.9$~MeV/s in TMBi and is taken into account in the final result as a systematic uncertainty. The mean energy absorbed per second was then found to be $124^{+4}_{-6}$~MeV/s in TMSi and $720^{+19}_{-38}$~MeV/s in TMBi.
	One should notice that the energy absorbed in TMBi is almost 6 times higher than in TMSi. This is the consequence of the high density and high photo-electric efficiency of TMBi.

\begin{figure}[htbp]
\centering 
\includegraphics[width=0.6\textwidth]{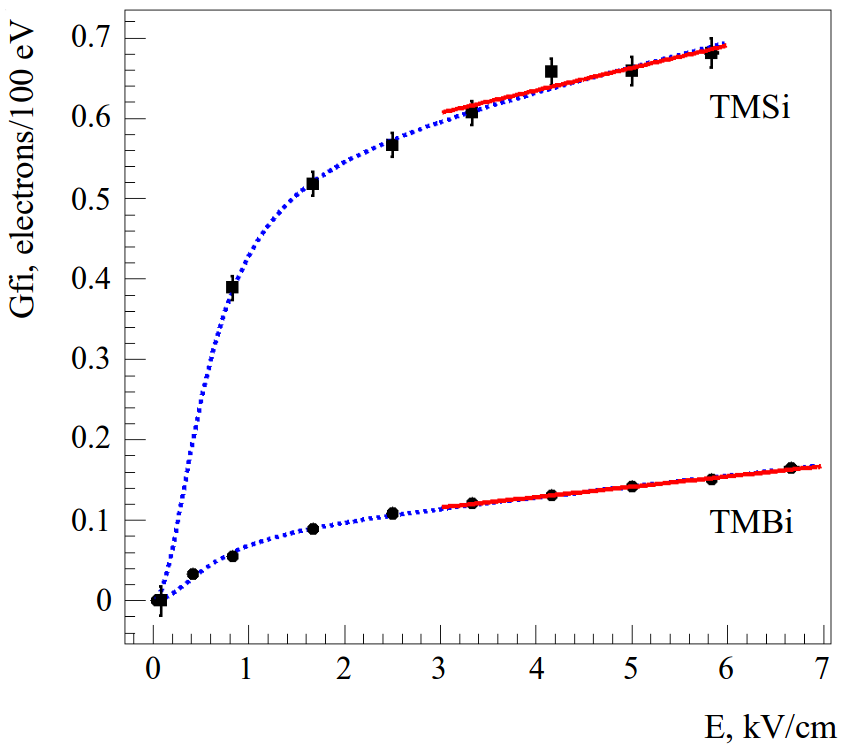}
\caption{Results of free ion yield as a function of electric field for TMSi (square markers) and TMBi (circle markers). The dashed blue line is the result of the model described by eq.~(\ref{eq_modele}). The solid red line corresponds to the linear term of Onsager theory (eq.~(\ref{eq_onsager})). Error bars represent measurement uncertainties.}
\label{fig_Gfi}
\end{figure}

	Finally, the free ion yield was estimated from eq.~(\ref{eq_gfi}). Figure~\ref{fig_Gfi} represents the free ion yield measured for TMSi and TMBi as a function of electric field strength. The measurement uncertainties are smaller in the case of TMBi, because we improved the device after measurements with TMSi by adding the power inverter, screened isolation transformer and local ground, which turned out to reduce significantly the measurement noise.

	We deduced the zero-field free ion yield and slope-to-intercept ratio from the linear term described by Onsager theory (eq.~\ref{eq_onsager}) to which we add a term describing collection efficiency as suggested in ref.~\cite{Par09}~:
\begin{equation}\label{eq_modele}
\Gfi(E) = \Gfi^0 (1+\alpha E) \left[ 1 + \frac{C (1+\alpha E)}{E^2} \right]^{-1}
\end{equation}
where $C$ is taken as a free parameter.
This simplified model allow us to describe the non-linear low-field behaviour of free ion yield. The results along with their respective uncertainties are summarized in table~\ref{table_result}. Constant $C$ was found to be $0.29 \pm 0.07$ for TMSi and $0.34 \pm 0.04$ for TMBi. 
We computed the systematic uncertainty by comparing the results of the linear model of Onsager theory (eq.~\ref{eq_onsager}) with the model described by eq.~\ref{eq_modele} and adding the systematic uncertainty of energy deposited in the liquids.

\begin{table}[htbp]
\centering
\caption{Zero-field free ion yields and slope-to-intercept ratios of TMSi and TMBi.}
\label{table_result}
\smallskip
\begin{tabular}{|ll|lll|lll|}
\hline
                   &      & \multicolumn{3}{c|}{$\Gfi^0$} & \multicolumn{3}{c|}{$\alpha$ (cm$\cdot$kV\minusone)}  \\
                   &      & value	& $\pm$stat		& $\pm$syst	& value	& $\pm$stat	& $\pm$syst  \\
\hline
Tetramethyl silane & TMSi & 0.53		& $\pm$0.03		& $\pm$0.03	& 0.05	& $\pm$0.02	& $\pm$0.0006 \\
Trimethyl bismuth  & TMBi & 0.083	& $\pm$0.003		& $\pm$0.005	& 0.15	& $\pm$0.02	& $\pm$0.007 \\
\hline
\end{tabular}
\end{table}

\section{Discussion}

\begin{figure}[htbp]
\centering 
\includegraphics[width=0.6\textwidth]{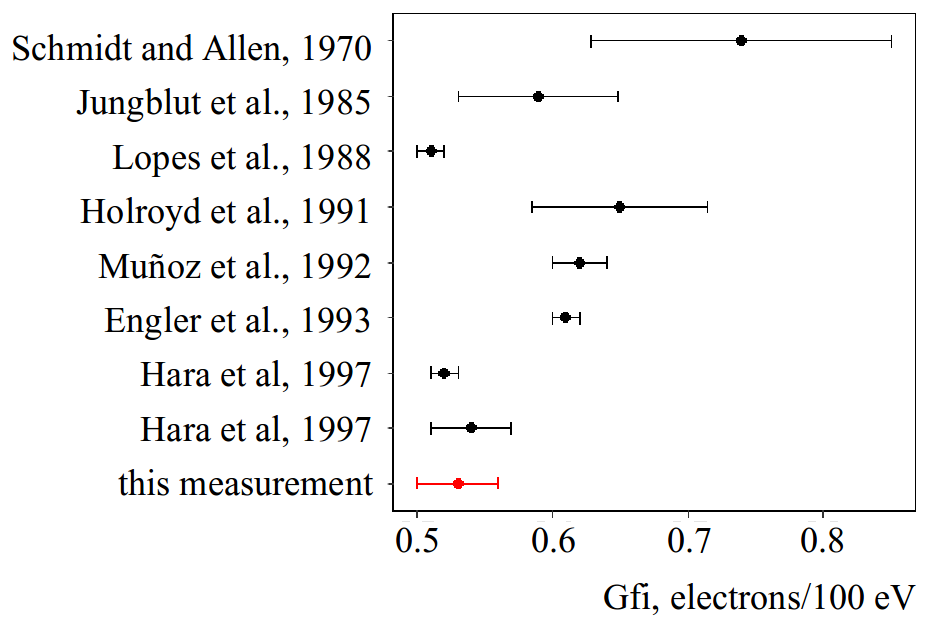}
\caption{Some values of TMSi free ion yield reported in literature in chronological order. Our measurement is represented in red.}
\label{fig_refsTMSi}
\end{figure}

	We investigated the values of zero-field free ion yield of TMSi in literature~\cite{Sch70,Jun85,Lop88,Hol91,Mun92,Eng93,Har98} as represented on figure~\ref{fig_refsTMSi}. The most precise measurements are not compatible with each other. For this reason, we did not represent the average of these reference values. Considering the uncertainty of the free ion yield measurement presented in this paper, one can conclude that this measurement is in agreement with literature.
	
	Free ion yield of TMBi is found to be about 7 times lower than TMSi. We expected a larger free ion yield, similar to those mentioned in table~\ref{table_gfi} for a collection of dielectric liquids. Holroyd et~al.~\cite{Hol91} showed that tetra-alkyl compounds of the same tetrahedral molecular structure have similar free ion yields regardless of the nature of the central atom. We assume that one of the reasons of the low free ion yield of TMBi can be found in its trigonal pyramidal molecular geometry.

		\paragraph{Quantum chemistry calculations.}	

	In order to investigate this assumption, we initiated relativistic density functional theory (DFT)-based calculations on TMSi and TMBi. We also ran these calculations with tetramethyl lead (TMPb) which have a heavy atom similarly to TMBi but the same tetrahedral structure than TMSi.
Within the context of DFT, Fukui functions are introduced as reactivity descriptors in order to identify the most reactive sites within a molecule~\cite{Bul03}. The Fukui function describes local changes occurring in the electronic density $\rho (\mathbf r)$ of the system due to changes in the total number of electrons $N$~\cite{Par84}.
Two definitions of Fukui functions depending on total electronic densities are given~:
\begin{subequations}
\begin{align}
f^+ &= \rho_{N+1}(\mathbf{r})-\rho_{N}(\mathbf{r})=\left( \frac{\partial\rho (\mathbf r)}{\partial N}\right)^+_{\nu(\mathbf r)} \\
f^- & =\rho_{N}(\mathbf{r})-\rho_{N-1}(\mathbf{r})=\left( \frac{\partial\rho (\mathbf r)}{\partial N}\right)^-_{\nu(\mathbf r)}
\end{align}
\end{subequations}
where $\rho_{N+1}(\mathbf{r})$, $\rho_{N}(\mathbf{r})$, $\rho_{N-1}(\mathbf{r})$ are the electronic densities at point $\mathbf{r}$ for the system with $N+1$, $N$ and $N-1$ electrons respectively, and $\nu(\mathbf r)$ an external potential. 
A dual descriptor has also been proposed by Morell et al.~\cite{Mor05} and is implemented as the difference between the Fukui plus and Fukui minus functions~:
\begin{equation}
f(\mathbf{r}) = f^{+} - f^{-}
\end{equation}

	Instead of using the Fukui function directly, it can be more convenient to condense their values around each atomic site into a single value that characterizes the atom in the molecule. Yang and Mortier~\cite{Yan86} proposed to approximate the Fukui function at the atom $k$ using the concept of condensed Fukui function as~:
\begin{subequations}
\begin{align}
f^+_k &= q_k^{N+1}-q_k^{N} \\
f^-_k &= q_k^{N}-q_k^{N-1}
\end{align}
\end{subequations}
where $q_k^{N+1}$, $q_k^{N}$ and $q_k^{N-1}$ are the atomic populations of the atom $k$ respectively in the $N+1$, $N$ and $N-1$ electron systems.
In these conditions, the function $f$ reflects the ability of a molecular site $k$ to accept  ($f^+_k$) or donate ($f^-_k$) electrons.
 
	The Fukui function was calculated using the Hirshfeld partitioning method \cite{Hir77}.  As noted by Roy et al. \cite{Roy99}, Hirshfeld partitioning has favourable practical properties for condensed reactivity indicators. For each molecule, a geometry optimization was performed for the neutral system with the TURBOMOLE program package \cite{Ahl89}, the PBE0 functional \cite{Per96}, the def2-TZVP basis set for all atoms and effective core potential for Bi (60 core electrons) \cite{Wei05}. In view of the large nucleus charge of bismuth, one should expect important effects of relativity on electronic properties. We have carried out the calculations in zero order regular approximation (ZORA) taking into account spin-orbit coupling with the ADF 2017 software package \cite{Vel01,ADF07}. The associated  all-electron ZORA/TZ2P basis sets were used for all atoms. The Fukui function was obtained using the meta-hybrid functional TPSSH \cite{Jia03,Sta03}.
The values calculated for $f^+_k$, $f^-_k$ and $f$ ($k=$ Si, Pb, Bi) are given in table \ref{f+}.

\begin{table}[htbp]
\centering
\caption{$f^+_k$, $f^-_k$ and $f$ values (in a.u.) for TMSi, TMPb and TMBi performing calculations that incorporate relativistic effects including spin-orbit interactions.}
\label{f+}
\smallskip
\begin{tabular}{|ll|c|c|c|}
\hline
&& $f^+_k$ & $f^-_k$ & Dual $f$ \\
\hline
Tetramethyl silane & TMSi & 0.065  & 0.103 & $-0.038$ \\
Tetramethyl lead & TMPb & 0.116 & 0.146 & $-0.030$ \\
Trimethyl bismuth & TMBi & 0.390 & 0.373 & $+0.017$ \\
\hline
\end{tabular}
\end{table}

	TMBi has an ability to donate an electron ($f^-_k$) larger than TMSi and TMPb, but also an ability to re-capture this electron ($f^+_k$) even more important. The dual descriptor reflects the global behaviour of molecules with TMSi and TMPb that donate electrons (negative dual~$f$) and TMBi that captures electrons (positive dual~$f$). The unexpected behaviour of TMBi is then more likely to be due to its trigonal pyramidal structure than to the presence of a heavy atom.

\begin{figure}[htbp]
\centering
\subfloat[TMSi]{\includegraphics[scale=0.1]{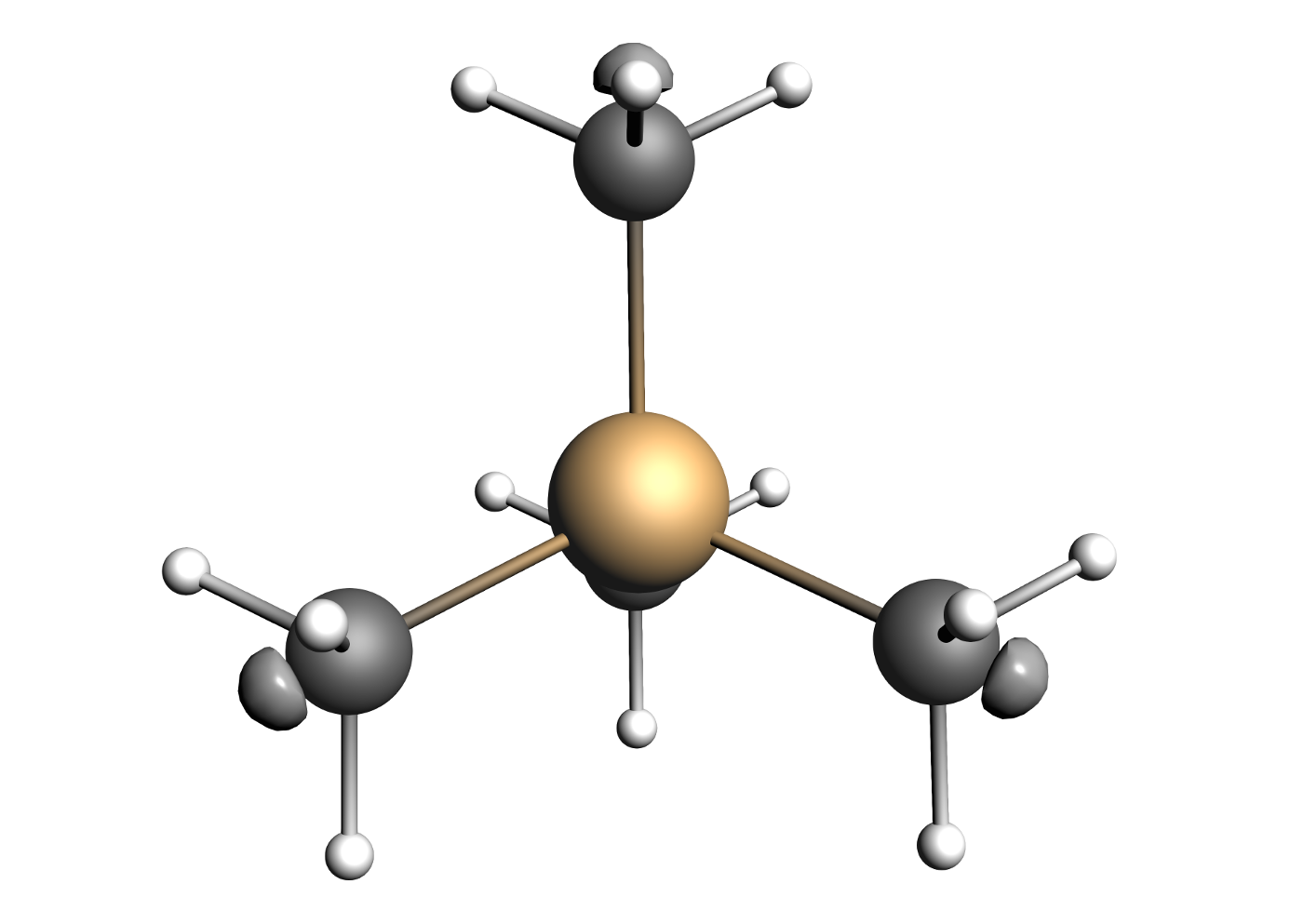}
\label{fig_f+TMSi}}
\hspace{2.1cm}
\subfloat[TMBi]{\includegraphics[scale=0.1]{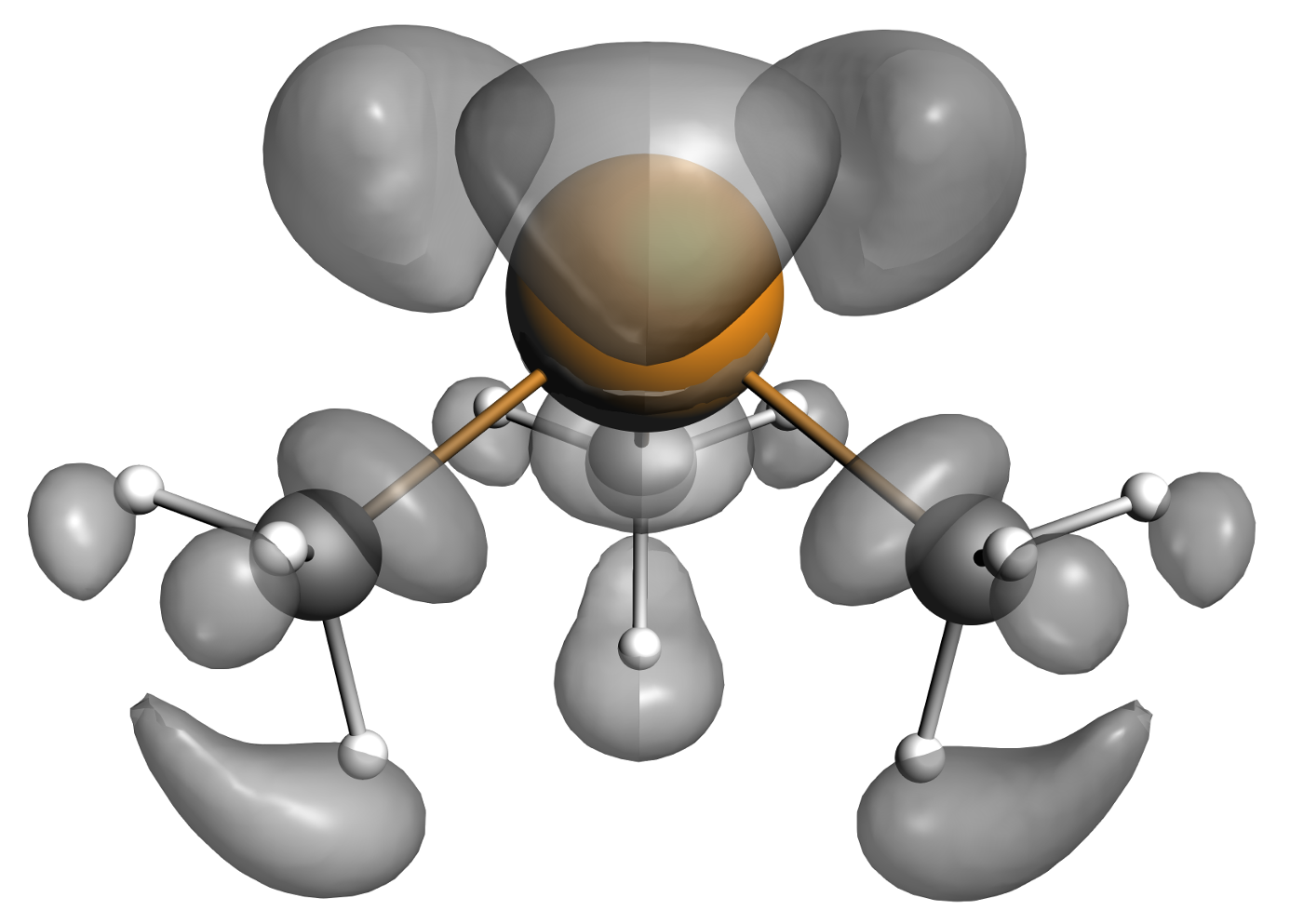}
\label{fig_f+TMBi}}
\\
\subfloat[TMSi]{\includegraphics[scale=0.1]{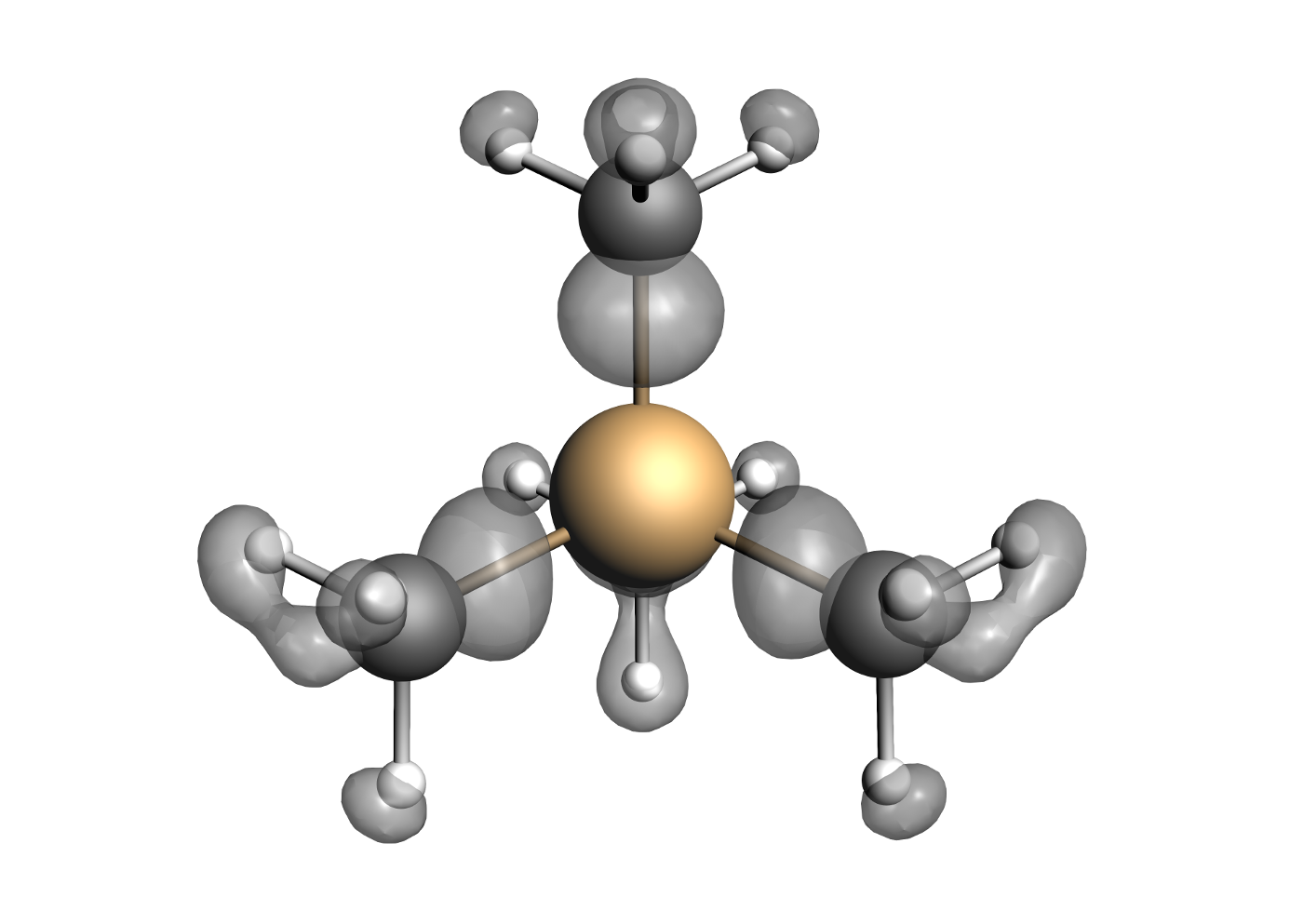}
\label{fig_f-TMSi}}
\hspace{2.1cm}
\subfloat[TMBi]{\includegraphics[scale=0.1]{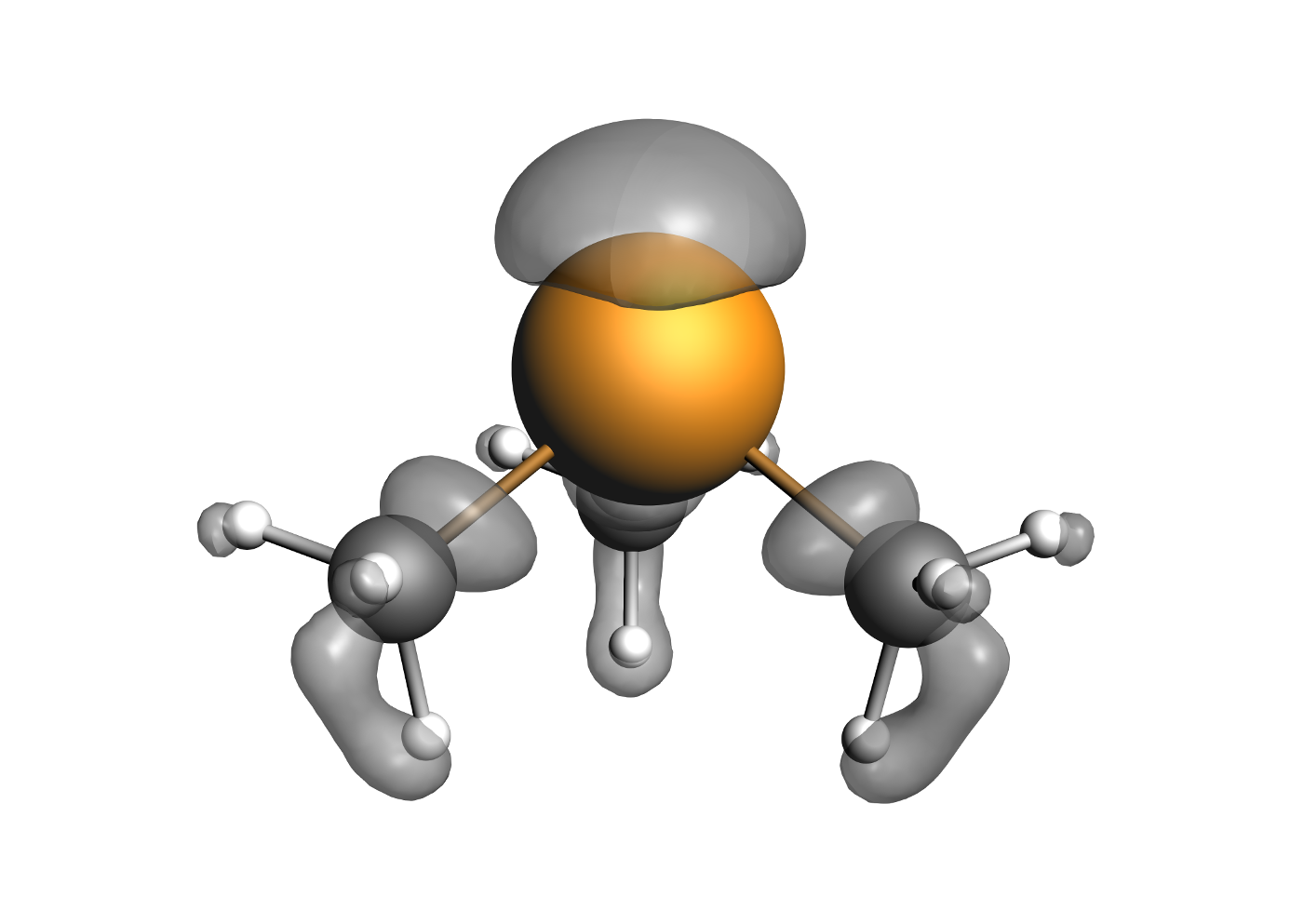}
\label{fig_f-TMBi}}
\caption{Isosurfaces of the Fukui function $f^+$= 0.0015 a.u. (a, b) and $f^-$= 0.005 a.u. (c, d) represented in grey color. Silicon and bismuth atoms are displayed respectively in light yellow and dark yellow.}
\label{fig_f_iso}
\end{figure}

	The descriptors can be analysed at the atomic level in the molecule. The isosurfaces of the condensed Fukui functions $f^+_k$ and $f^-_k$ are plotted in figure~\ref{fig_f_iso} for TMSi and TMBi. The condensed Fukui function isosurfaces bulge outward on bismuth atom indicating that Bi is a favourable site to add or loose electrons. The opposite occurs on TMSi molecule where the electron donation or capture is  controlled in both cases by methyl groups.

	These results therefore tend to demonstrate that electrons are more easily released from TMBi than TMSi. However, the higher ability of TMBi to capture electrons compensates this effect so that electrons are more likely to get attached to another molecule which then becomes an anion. Because of its high density, the electron attachment to a TMBi molecule may occur near the parent cation. A recombination between this anion and the parent cation is then likely to occur. This phenomenon is analogous to initial recombination and could explain the low value measured for $\Gfi^0$ in our experiment with TMBi.
	
	In light of the above calculations, TMPb might be an interesting alternative detection medium. The high atomic number of lead results in a high photoelectric cross section comparable to that of TMBi, and the free ion yield of TMPb is likely to be close to the one of TMSi. However, further investigation is needed : liquid TMPb has a lower density than TMBi, its free ion yield still has to be measured and its optical properties have not yet been studied. Finally, its toxicity~\cite{Cre61} would also require appropriate handling protocols.

		\paragraph{Onsager theory predictions.}
	
	The slope-to-intercept ratio $\alpha$ for TMSi given by Onsager equation~(\ref{eq_alpha}) is expected to be $0.0588 \pm 0.0008$~cm$\cdot$kV\minusone\ considering a temperature of $20 \pm 2^\circ$C. As shown on table~\ref{table_result}, our measurement corroborates Onsager theoretical prediction.
	In the case of TMBi, the dielectric constant $\epsilon_r$ can be estimated from the refractive index~$n$ published in a previous paper~\cite{Ram16}~: $n = 1.58 \pm 0.06$ for 700 nm in wavelength, leading to~: $\epsilon_r = n^2 = 2.5 \pm 0.2$. We thus expected a slope-to-intercept ratio $\alpha$ of $0.045 \pm 0.004$~cm$\cdot$kV\minusone\ for TMBi. The measured value of $\alpha$ is however almost 4 times higher than this theoretical prediction. This result shows a breakdown of the Onsager theory for this very specific liquid which differs from other dielectric liquids in terms of density, atomic number and trigonal pyramidal shape.

		\paragraph{Consequences.}
	
	In terms of high-energy photon detection, a lower free ion yield implies a lower charge signal amplitude, leading to a degradation of energy resolution. For example, with an operating electric field of 7~kV$\cdot$cm\minusone, the amplitude of the signal produced by a 500~keV photon would be $0.13$~fC in TMBi. The statistical limitation on the energy resolution, neglecting readout electronics, is then : $\sqrt{N}/N = 3.5$~\%. In comparison, in liquid Xenon, the free ion yield for a field of 7~kV$\cdot$cm\minusone\ reaches approximately 6.1~electrons per 100~eV of absorbed energy~\cite{Sch02}, resulting in an amplitude charge signal of 4.9~fC for a 500~keV photon and an energy resolution limited to 0.6~\%.
	
	However, if the signal-to-noise ratio is sufficient, the charge signal still allows accurate positioning of photon interactions in TMBi. IDeF-X ASICs will be used as charge amplifier in the final pixelated detection device. According to ref.~\cite{Gev12} and considering a detector capacitance of 0.1~pF and a dark current $<1$~pA, the  equivalent noise charge is estimated at 35~electrons RMS. The signal produced by a 500~keV photon in TMBi is approximately 825~electrons for an operating voltage of 7~kV$\cdot$cm\minusone. The signal-to-noise ratio is then greater than 20, which allows the charge signal to be used to locate photon interactions in TMBi.

	Finally, the CaLIPSO detector was designed for high resolution brain PET imaging. PET system electronics apply an energy acceptance criterion to reject part of the scatter coincidences. In this context, a degradation of energy resolution entails a reduced ability to discriminate scatter events and a deterioration of image contrast.
	However, the signal induced in the CaLIPSO detector is triggered by a Cherenkov radiation which is detectable only for electrons with an energy superior to 250~keV~\cite{Ram16}. Part of scatter events will thus be discriminated and the deterioration of image contrast will be mitigated.
	We now need a full scanner Monte Carlo simulation to assess the consequences for brain PET imaging and to quantify the image quality achievable with an ionization chamber filled with TMBi.

	\subsection{Foreseen developments}
	
	We are now working on the measurement of charge pulses produced by the interaction of a single 511~keV~photon. This will allow us to assess the impact of the electron capture ability of TMBi on charge signals. Would charge collection in the detector be complete, we will be able to confirm our measurement of TMBi free ion yield by a second method. In parallel, we are developing a Frisch-grid ionization chamber that will allow us to measure the mobility of free electrons and the energy resolution for 511~keV~photons.

\section{Conclusion}

	We developed an accurate measuring device of radiation induced currents in chemically reactive dielectric liquids. We duplicated a purification system based on molecular sieves for TMSi and adapted it to TMBi. We measured the free ion yield of both liquids. The measured zero-field free ion yield of TMBi is lower than similar dielectric liquids and slope-to-intercept ratio of $\Gfi(E)$ is higher than predicted by Onsager theory.
In order to understand these unexpected results, we ran quantum chemistry calculations for both molecules. The ability of TMBi molecule to capture electrons implies an additional trapping mechanism for ionization electrons near their parent cations that might explain the low $\Gfi^0$ value.
Future measurements of single interaction charge pulses on liquids TMSi and TMBi will allow us to test these assumptions quantitatively.

\acknowledgments

We thank C. Weinheimer (Institut f\"ur Kernphysik, Germany), K.P. Sch\"afers (European Institute for Molecular Imaging, Germany) and their respective work teams for our emerging collaboration on the CaLIPSO project and its perspectives. We are grateful to J-Ph.~Renault and D.~Doizi for fruitful discussions and to the engineering teams of IRFU / DEDIP for technical support.


\end{document}